\documentclass[twocolumn,nofootinbib,aps,showpacs,preprintnumbers,showkeys,natbib]{revtex4}
\usepackage{graphics}
\usepackage{amsfonts}
\usepackage{latexsym}

\begin{document}

\preprint{UB-ECM-PF 05/09, IRB-TH-17/05}

%\begin{flushright}
%{\normalsize IRB-TH-4/99 \\
%July, 1999.}
%\end{flushright}
%\vspace{1cm}
%\thispagestyle{empty}
%\title{
%\vspace{-2cm}
%\begin{flushright}
%{\normalsize IRB-TH-12/01 \\
%\vspace{-0.5cm}
%October 2001}
%\end{flushright}
%\vspace{1 cm}
\title{Dark energy transition between quintessence and phantom regimes \\- an equation of state analysis}

%\vspace{1cm}
%\author{A. Babi\'c\thanks{E-mail: ababic@thphys.irb.hr}, B. Guberina\thanks{E-mail: guberina@thphys.irb.hr},
%Tel: +385-1-4680234; fax: +385-1-4680223;
% R. Horvat\thanks{E-mail: horvat@lei3.irb.hr},
% H. \v Stefan\v ci\'c\thanks{E-mail: shrvoje@thphys.irb.hr}
%}
\author{Hrvoje \v Stefan\v ci\'c \footnote{On leave of absence from the  Theoretical Physics Division, Rudjer Bo\v{s}kovi\'{c} Institute, Zagreb, Croatia.}}
\email{stefancic@ecm.ub.es}

%\vspace{2cm}
%\date{
%\centering
%Theoretical Physics Division,
\affiliation{Departament d' Estructura i Constituents de la Mat\`{e}ria, Universitat de Barcelona \\
Av. Diagonal 647, 08028 Barcelona, Catalonia, Spain }

%\institute{
%  Theoretical Physics Division, Ru\dj er Bo\v{s}kovi\'{c} Institute,
%   P. O. Box 1016, HR-10001 Zagreb, Croatia}

\begin{abstract}
The dark energy transition between quintessence ($w>-1$) and phantom ($w<-1$)  regimes (the crossing of the cosmological constant boundary) is studied using the dark energy equation of state. Models characterized by this type of transition are explicitly constructed and their equation of state is found to be {\em implicitly} defined. The behavior of the more general models with the implicitly defined equation of state, obtained by the generalization of the explicitly constructed models, is studied to gain insight into the necessary conditions for the occurrence of the transition, as well as to investigate the mechanism behind the transition. It is found that the parameters of the generalized models need to satisfy special conditions for the transition to happen and that the mechanism behind the transition is the cancellation of the contribution of the cosmological constant boundary. The aspects of the behavior of the generalized models which are not related to the transition are briefly discussed and the role of the implicitly defined dark energy equation of state in the description of the dark energy evolution is emphasized.
\end{abstract}

%\vspace{1cm}

\noindent
%\vspace{1cm}

\maketitle

One of the most important aspects of the present universe is its accelerated expansion. Numerous and complementary cosmological observations such as supernovae of the type Ia (SNIa) \cite{SNIa}, 
the anisotropies of the cosmic microwave background radiation (CMBR) \cite{CMB}, large scale structure (LSS) \cite{LSS}, and others seem to reconfirm this characteristic of the small-redshift evolution of the universe with the arrival of each new set of cosmological data. Although the present acceleration of the universe expansion is well established observationally, the nature of the cause of the cosmic acceleration is much more uncertain.
A large number of models assume the existence of the component of the universe with the negative pressure, named {\em dark energy} \cite{rev}, which at late times dominates the total energy density of the universe and accelerates its expansion. Models of dark energy differ with respect to the size of the parameter $w=p_{d}/\rho_{d}$ ($p_{d}$ and $\rho_{d}$ are the pressure and energy density of dark energy, respectively \footnote{Since only the dark energy component of the universe will be discussed furtheron, in the remainder of the paper the subscript $d$ will be dropped.}) of their equation of state (EOS) as well as the variation of the parameter $w$ with redshift or cosmic time.
The {\em cosmological constant} (CC), with $w=-1$, occupies the central place among the dark energy models, both in theoretical considerations and in data analysis \cite{cc}. The conceptual difficulties in the understanding of the measured size of the CC and its relation to other cosmological parameters motivated the study of the dynamical models of CC (such as renormalization group running models of CC and other relevant cosmological parameters \cite{Sola,mi,ShSoSt,ReuterGen}) and dynamical models of dark energy in general. The models of dark energy with $w>-1$ comprise {\em quintessence} \cite{Q}, {\em k-essence} \cite{k}, and {\em Chaplygin gas} \cite{Chaplygin}, among others. On the other side of the CC boundary, characterized by $w<-1$, lie {\em phantom} models of dark energy, recently intensively studied in \cite{Caldwell,phantom,phantomja,phantom2,bigrip}(for an intersting explanation of superaccelerated expansion without phantom components, see \cite{onemli}).
The second important direction in the study of the present acceleration of the universe explains it in terms of the modifications of the gravitational interaction at cosmological scales \cite{modgrav}. 
%Finally, some recently proposed models \cite{a} study the role of the superhorizon perturbations in understanding the present acceleration of the universe (see also \cite{a}). %>>> <<<

Although there exist different approaches to the problem of the acceleration of the universe, the concept of dark energy is especially useful and most widely used. Namely, the dynamics of the universe in the approaches which do not assume the existence of dark energy can still be described in terms of dark energy, which in this case becomes an effective description. In this paper we model specific aspects of the expansion of the universe in the framework of dark energy, which may be either a fundamental cosmic component or an effective description of some alternative mechanism causing the acceleration of the universe.

The formalism of this paper is formulated in terms of the dark energy EOS. The usual formulation of the dark energy EOS is given in terms of $p=p(\rho)$ where $p$ is some analytic function of $\rho$. This form of dark energy EOS may be inappropriate for the study of some effects related to dark energy. Therefore we adopt a more general approach in which we consider {\em parametrically defined dark energy EOS} as a pair of functions $(\rho(t),p(t))$ or $(\rho(a),p(a))$ where $t$ is the cosmic time and $a$ is the scale factor of the universe. The parameter of dark energy EOS can also be defined in a straightforward manner. This definition is general enough to describe the behavior of large classes of dark energy, irrespectively of their internal degrees of freedom, i.e. their more detailed formulation.  

Within the description of dark energy in terms of the EOS as defined in the preceding paragraph, it is possible to address a problem which has recently drawn a lot of attention of researchers in both observational \cite{obser} and theoretical cosmology \cite{hu,guo,wei,quintom,vikman,mingzheli,calddor}: a possible dark energy {\em transition between quintessence and phantom regimes}. The analyses of various cosmological data sets \cite{obser} mildly favor the evolution of the dark energy parameter of EOS from $w>-1$ to $w<-1$ at small redshift. It is important to stress that there is no decisive indication for this type of transition from the data and there are other models (especially the benchmark $\mathrm \Lambda CDM$ model) which are consistent with the presently available cosmological data. However, should the future, more precise and abundant cosmological data confirm the existence of the aforementioned type of transition, the theoretical understanding of this phenomenon will become necessary. The transition from quintessence to phantom regime of the behavior of dark energy is also theoretically favorable. Namely, since the energy density of phantom energy grows with the expansion of the universe, the coincidence problem is for phantom energy even more serious than for the CC. However, if at early times the dark energy has the quintessence character ($w>-1$) and only at some (relatively) small redshift transits to phantom, the coincidence problem is substantially alleviated. The quintessence-like behavior up to the transition makes possible the development of models which incorporate the feature that the quintessence energy density may track energy densities of other components of the universe. Moreover, the theoretical description of the CC boundary transition unifies aspects of many different classes of dark energy models (such as quintessence, phantom and CC models) and represents a challenge of substantial theoretical importance. A number of theoretical studies of the crossing of CC boundary have been undertaken so far. The approach of \cite{hu,guo} (see also \cite{wei}) models the dark energy in terms of two fields, of which one is of quintessence and the other of phantom type. An interesting phenomenological model of oscillatory parameter $w$, named {\em quintom} \cite{quintom}, exhibits very interesting features in the evolution of the universe. In the {k-essence} single scalar field models it has been shown that the evolution of quintessence to phantom is unlikely \cite{vikman}, but the addition of higher-derivative kinetic terms in the single scalar field action can reproduce the transition between the quintessence and phantom regimes of dark energy \cite{mingzheli}. A possible imprint of this transition on cosmological perturbations, along with other aspects of the transition, was studied in \cite{calddor}.            

In this paper we search for the dark energy EOS (understood in the sense defined above) which allows for the transition to occur and we study the conditions for the transition within the generalized classes of dark energy models. Some interesting results on the CC boundary crossing using the dark energy EOS have been obtained in \cite{odin1}. For the dark energy EOS in which $p$ is defined explicitly in terms of $\rho$
\begin{equation}
\label{eq:fodro}
p = -\rho - f(\rho) \, ,
\end{equation}    
the parameter of the dark energy EOS is given by 
\begin{equation}
\label{eq:wdef}
w = -1 -f(\rho)/\rho \, .
\end{equation} 
Let us assume that the transition between the quintessence and phantom regimes happens at the dark energy density $\rho_{*}$. Then at the point of transition the function $f$ must vanish, $f(\rho_{*})=0$, and needs to change sign at the same value of the dark energy density.  The equation of the energy-momentum tensor conservation 
\begin{equation}
\label{eq:bianchi}
d\rho + 3 (\rho+p) \frac{d a}{a} = 0\, 
\end{equation}
yields the following expression for the evolution of the dark energy density towards the point of transition
%
%, $\rho_{1} < \rho_{*} < \rho_{2}$ 
%
\begin{equation}
\label{eq:integr}
\int_{\rho}^{\rho_{*}} \frac{du}{f(u)} = 3 \ln \left( \frac{a_{*}}{a} \right) \, .
\end{equation}
Since we wish to describe the crossing of the CC boundary at finite values of the scale factor (i.e. redshifts), the function $1/f(u)$ must be integrable in the considered interval of $\rho$. The specified requirements put considerable constraints on the form of allowable functions $f$. One of possibilities is that the function $f$ is multivalued near the point of transition $\rho_{*}$ \cite{odin1}. An acceptable possibility is also  $f(\rho) \sim (\rho-\rho_{*})^s$, with $0 < s < 1$ \cite{odin1}. 

In our concrete study of the transition phenomenon we start from a simple expression for the evolution of the dark energy density
\begin{equation}
\label{eq:denmod1}
\rho = C_{1} \left( \frac{a}{a_{0}} \right)^{-3(1+\gamma)} + 
C_{2} \left( \frac{a}{a_{0}} \right)^{-3(1+\eta)} \, .
\end{equation}
In the expression given above $\gamma>-1$ and $\eta<-1$, whereas $C_{1}$ and $C_{2}$ are positive constants.
This expression resembles the sum of contributions of two independent cosmological fluids with constant parameters of EOS $\gamma$ and $\eta$. However, we consider (\ref{eq:denmod1}) to be the energy density of a $single$ cosmological component and we further investigate its EOS. From (\ref{eq:bianchi}) the expression for the dark energy pressure can be obtained in a straightforward manner:  
\begin{equation}
\label{eq:presmod1}
p =  \gamma C_{1} \left( \frac{a}{a_{0}} \right)^{-3(1+\gamma)} + 
 \eta C_{2} \left( \frac{a}{a_{0}} \right)^{-3(1+\eta)} \, .
\end{equation}
\begin{figure}
\centerline{\resizebox{0.45\textwidth}{!}{\includegraphics{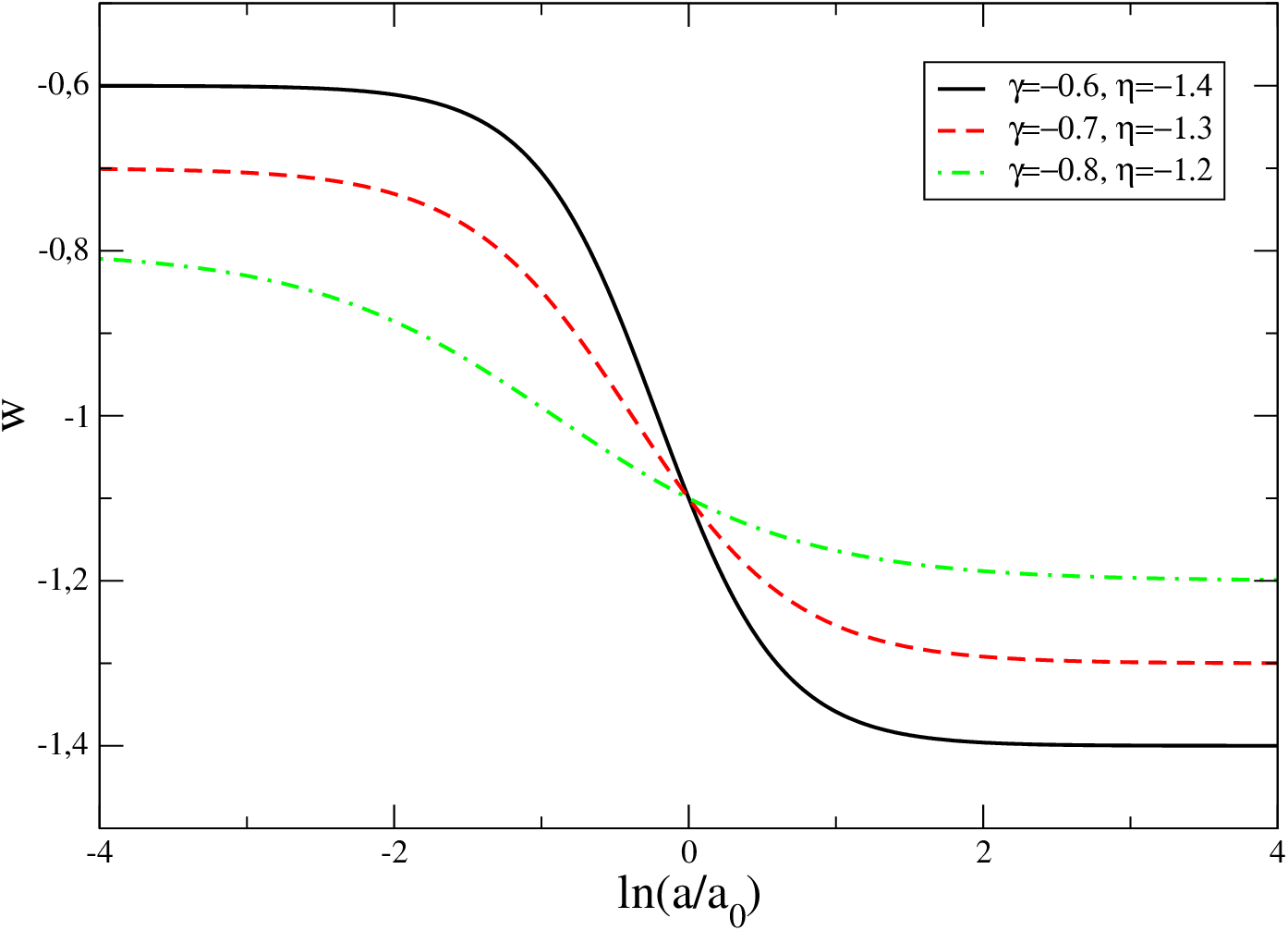}}}
\caption{\label{fig:mod1} The dependence of the dark energy parameter of EOS $w$ given by (\ref{eq:wformod1}) on the scale factor $a$ for the present value $w_{0}=-1.1$ and three sets of parameters $\gamma$ and $\eta$.  
}
\end{figure}
The parameter of EOS acquires the form 
\begin{equation}
\label{eq:wformod1}
w=\frac{\gamma + \eta \frac{\gamma - w_{0}}{w_{0} - \eta} \left( \frac{a}{a_{0}} 
\right)^{3(\gamma - \eta)}}{1 + \frac{\gamma - w_{0}}{w_{0} - \eta} \left( \frac{a}{a_{0}} 
\right)^{3(\gamma - \eta)}} \, ,
\end{equation}
where $w_{0}$ denotes the present value of the parameter of dark energy EOS. From the functional form (\ref{eq:wformod1}) it is clear that for small values of the scale factor the parameter $w$ tends to $\gamma$ while for large values of the scale factor it tends to $\eta$ and that it transits from one asymptotic value to the other at some finite value of the scale factor. The dependence of $w$ on $a/a_{0}$ is given in Fig. \ref{fig:mod1}.
From equations (\ref{eq:denmod1}) and (\ref{eq:wformod1}) it can be seen that the constants $C_{1}$ and $C_{2}$ can be expressed in terms of the present values of the dark energy density $\rho_{0}$ and the parameter of EOS $w_{0}$. 
Since (\ref{eq:denmod1}) and (\ref{eq:presmod1}) are essentially linear combinations of power terms in $a/a_{0}$, it is possible to eliminate the scale factor in the following way
\begin{equation}
\label{eq:eosmod1}
\left( \frac{a}{a_{0}} \right)^{-3} = \left( 
\frac{\gamma \rho - p}{(\gamma-\eta) C_{2}} \right)^{1/(1+\eta)}
= \left( \frac{p -\eta \rho}{(\gamma-\eta) C_{1}} \right)^{1/(1+\gamma)} \, ,
\end{equation}
which finally leads to the EOS for the dark energy model (\ref{eq:denmod1})
\begin{equation}
\label{eq:eosdetmod1}
\frac{p -\eta \rho}{(\gamma-\eta) C_{1}} =
\left( \frac{\gamma \rho - p}{(\gamma-\eta) C_{2}} \right)^{(1+\gamma)/(1+\eta)}
\, .
\end{equation}

In the procedure described above we have constructed the EOS for a well motivated and simple example. The obtained
EOS has some features which distinguish it from the dark energy equations of state considered in the literature so far.
The most important one is that, for general values $\gamma>-1$ and $\eta<-1$, the EOS is {\em implicitly} defined and generally it is not possible to obtain a closed form analytic expression $p=p(\rho)$. 

The features of the dark energy EOS constructed above motivate us to consider more general implicitly defined equations of state. A natural generalization of EOS (\ref{eq:eosdetmod1}) is given by 
\begin{equation}
\label{eq:eosgenmod1}
A \rho + B p = (C \rho + D p)^{\alpha} \, ,
\end{equation}
where $A$, $B$, $C$, and $D$ are real coefficients and we consider the case $\alpha \neq 1$ since for $\alpha=1$ (\ref{eq:eosgenmod1}) becomes an EOS with a constant parameter $w$. Since (\ref{eq:eosdetmod1}) is a special case of (\ref{eq:eosgenmod1}), we know that the general model necessarily includes some parameter sets which describe the dark energy transiting from the quintessence to the phantom regime. A natural question before the detailed considerations of the model (\ref{eq:eosgenmod1}) is whether all parameter sets lead to the transition. The answer to this question is negative. Namely, for $D=0$ the model, with a suitable choice of parameters $A$, $B$, and $C$, acquires the form  $p = -\rho - K \rho^{\delta}$ which was studied in detail in \cite{PRDja} and \cite{odin1}. This model exhibits many interesting effects, but its parameter $w$ remains either on the quintessence or the phantom side during the evolution of the universe. Therefore, the model has some parts of the parameter space in which transition occurs and some in which it does not occur. Since the model defined by (\ref{eq:eosgenmod1}) contains a number of parameters, in our considerations we shall focus on those parameter values which allow the study of the quintessence-phantom transition. From (\ref{eq:eosgenmod1}) it is straightforward to obtain the expression for $\rho$ in terms of the EOS parameter $w$:
\begin{equation}
\label{eq:rhoexpr1}
\rho = \frac{(C+ D w)^{\alpha/(1-\alpha)}}{(A + B w)^{1/(1-\alpha)}} \, .
\end{equation}
We further introduce the parameters $E=A/B$ and $F=C/D$. 
%We shall in general be interested in finite values of these parameters although the situations when they become infinite will also be briefly commented. 
Combining the equation (\ref{eq:rhoexpr1}) with the evolution law (\ref{eq:bianchi}) we obtain the following equation for the evolution of the parameter $w$ with the scale factor $a$:
\begin{equation}
\label{eq:evolw1}
\left( \frac{\alpha}{(F+w)(1+w)}-\frac{1}{(E+w)(1+w)} \right) dw = 3(\alpha - 1)
\frac{da}{a} \, .
\end{equation}
The functional form of the solution of this differential equation depends on whether some of the coefficients $E$ and $F$ equals $1$ or not. For the case when $E \neq 1$ and $F \neq 1$, the solution of (\ref{eq:evolw1}) acquires the form
\begin{eqnarray}
\label{eq:solw}
&&\left| \frac{w+F}{w_{0}+F} \right|^{\alpha/(1-F)} 
\left| \frac{w+E}{w_{0}+E} \right|^{-1/(1-E)} \nonumber \\
&\times&\left| \frac{1+w}{1+w_{0}} \right|^{1/(1-E)-\alpha/(1-F)}
= \left( \frac{a}{a_{0}} \right)^{3(\alpha-1)} \, . 
\end{eqnarray} 
This solution reveals interesting aspects of the evolution of $w$ with the expansion of the universe. The values $-1$, $-E$, and $-F$ of the parameter $w$ can generally be reached only for the asymptotic values of the scale factor $a$, i.e. $a \rightarrow 0$ or $a \rightarrow \infty$, depending on the concrete numerical values of parameters $\alpha$, $E$, and $F$. The values $-1$, $-E$ and $-F$ therefore represent the boundaries that cannot be crossed, {\em whenever these boundaries exist in the problem}. The evolution of the parameter $w$ with the scale of the universe $a$ is always confined into one of the intervals obtained by the division of the $w$ axis with the points $-1$, $-E$, and $-F$. The interval within which the parameter $w$ evolves is determined by the choice of the present-time value $w_{0}$. 
For the choice of parameters as in this case, $E \neq 1$ and $F \neq 1$,  these boundaries can be removed by the choice of the parameter $\alpha$. The boundary at $w=-F$ is removed when $\alpha = 0$ and the boundary at $w=-E$ is removed when $\alpha \rightarrow \pm \infty$. The most interesting case is the possibility of removing the boundary at $w=-1$. It is achieved for $\alpha_{\rm cross} = (1-F)/(1-E)$. This choice gives the transition from $w  >-1$ to $w<-1$, i.e. the crossing of the CC boundary. The equation (\ref{eq:solw}) then acquires the form
\begin{equation}
\label{eq:solwcross}
\left| \frac{w+F}{w_{0}+F} \right| 
\left| \frac{w+E}{w_{0}+E} \right|^{-1}
= \left( \frac{a}{a_{0}} \right)^{3(E-F)} \, . 
\end{equation}
The parameter $w$ smoothly varies between the asymptotic values $w=-E$ and $w=-F$.       
In the case when $E = 1$ and $F \neq 1$ the solution of (\ref{eq:evolw1}) is
\begin{eqnarray}
\label{eq:solwe1}
&&\left| \frac{w+F}{w_{0}+F} \right|^{\alpha/(1-F)} 
\left| \frac{1+w}{1+w_{0}} \right|^{-\alpha/(1-F)} \nonumber \\
&\times& e^{1/(1+w)-1/(1+w_{0})}
= \left( \frac{a}{a_{0}} \right)^{3(\alpha-1)} \, .
\end{eqnarray}
This solution demonstrates that the choice of $\alpha$ that could remove the CC boundary does not exist. Therefore, in this case we cannot obtain the transition between quintessence and phantom regimes.
For $E \neq 1$ and  $F=1$ the solution has the form
\begin{eqnarray}
\label{eq:solwf1}
&& e^{-\alpha(1/(1+w)-1/(1+w_{0}))} 
\left| \frac{w+E}{w_{0}+E} \right|^{-1/(1-E)} \nonumber \\
&\times& \left| \frac{1+w}{1+w_{0}} \right|^{1/(1-E)}
= \left( \frac{a}{a_{0}} \right)^{3(\alpha-1)} \, , 
\end{eqnarray}
and as in the preceding case the crossing of the CC boundary cannot be achieved by a suitable choice of $\alpha$.
Finally, when both coefficients equal 1, i.e. $E=1$ and $F=1$, we obtain the following solution:
\begin{equation}
\label{eq:E1F1}
w = -1 + \frac{1+w_{0}}{1-3 (1+w_{0}) \ln (a/a_{0})} \, .
\end{equation}
This solution does not depend on $\alpha$ and cannot describe the transition between the quintessence and phantom regimes as well.
The relations (\ref{eq:solw}) to (\ref{eq:E1F1}) show that the evolution of the parameter $w$ for the generalized model (\ref{eq:eosgenmod1}) can be expressed in the closed form and that the generalized model (\ref{eq:eosgenmod1}) is highly analytically tractable. 
%Since the crossing of the CC boundary can be realized only for the $E \neq 1$ and $F \neq 1$ case, we further concentrate on the study of the model behavior for these values of parameters.
%$\alpha_{\rm cross} = (1-F)/(1-E)$.

The equation (\ref{eq:evolw1}) can be rearranged to provide some further insight into the mechanism of the CC boundary crossing. Namely, it can be written as  
\begin{equation}
\label{eq:wstar}
\frac{w+\frac{\alpha E - F}{\alpha - 1}}{(F+w)(E+w)(1+w)} dw = 3 \frac{da}{a} \, .
\end{equation}
From the equation given above it is evident that, apart from zeros of the denominator of the expression on the left-hand side of the equation at $w=-1$, $w=-E$, and $w=-F$, the numerator of the same expression has a zero at  
$w_{*} = -(\alpha E - F)/(\alpha - 1)$. This form of the equation governing the evolution of $w$ casts additional light on the possible removal of the boundaries associated with the zeros of the denominator of the expression on the left hand side. Namely, when $w_{*}$ equals one of the zeros of the denominator, there is the cancellation of the terms in denominator and numerator which removes this zero from the denominator and the boundary associated with this zero of the denominator is not present in the solution for $w$. In this way we can see that $w_{*}=-F$ for $\alpha=0$, $w_{*}=-E$ for $\alpha \rightarrow \pm \infty$, and $w_{*}=-1$ for $\alpha_{\mathrm cross}=(1-F)/(1-E)$. In this way we confirm the results obtained by the analysis of the solution (\ref{eq:solw}). The conclusion is that the mechanism behind the crossing of any of the boundaries $w=-1$, $w=-E$, and $w=-F$ is the cancellation of terms in the numerator and denominator of the expression on the left-hand side of (\ref{eq:wstar}). 

The equation (\ref{eq:wstar}), apart from the described insight into the crossing mechanism, provides additional information on the behavior of the general model (\ref{eq:eosgenmod1}) which cannot be easily obtained from the solutions (\ref{eq:solw}) to (\ref{eq:E1F1}). Namely, from (\ref{eq:wstar}) it is possible to obtain the solution in the very vicinity of $w_{*}$ which has the form
\begin{equation}
\label{eq:aroundwstar}
a = a_{*} e^{(w-w_{*})^2/(6 L)} \, ,
\end{equation}
where 
\begin{equation}
\label{eq:L}
L = \frac{\alpha}{(\alpha-1)^3}(F-E)^2 (\alpha (1-E) - (1-F)) \, .
\end{equation}

\begin{figure}
\centerline{\resizebox{0.45\textwidth}{!}{\includegraphics{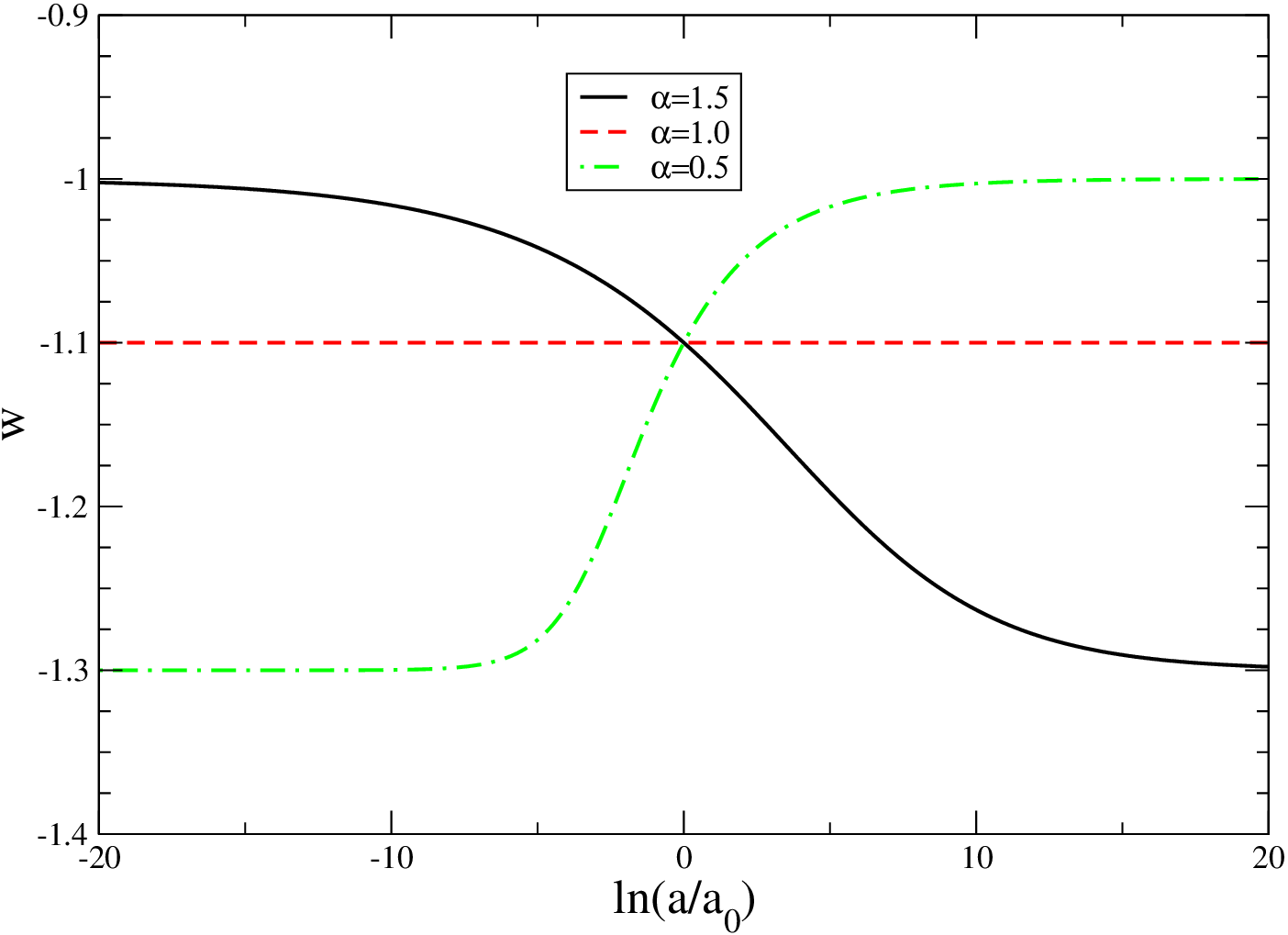}}}
\caption{\label{fig:veciod0} The scaling of the parameter $w$ with $a$ for the model (\ref{eq:eosgenmod1}) for three typical values $\alpha=1.5$, $\alpha=1$, and $\alpha=0.5$ in the interval $\alpha > 0$. The parameter values used are $w_{0}=-1.1$, $E=0.6$ and $F=1.3$. The values of the parameter $w_{*}$ are $0.8$, $\infty$, $-2$ for $\alpha=1.5, 1, 0.5$ respectivelly, and are outside the interval $(-F,-1)$.   
}
\end{figure}

 Here $a_{*}$ is the value of the scale factor which corresponds to the EOS parameter value $w=w_{*}$. The expression (\ref{eq:aroundwstar}) is valid when $w_{*}$ is different from $-1$, $-E$, or $-F$, i.e. when no cancellation mechanism is at work. Depending on whether $L$ is positive or negative, the expression (\ref{eq:aroundwstar}) describes the evolution of the universe with a minimal or a maximal value of the scale factor $a_{*}$ which is reached when $w=w_{*}$. Therefore, $w_{*}$ is, apart from $-1$, $-E$ and $-F$, the fourth typical value for $w$ which determines the behavior of the model. The principal difference compared to the first three values is that $w_{*}$ depends on $\alpha$ and, for fixed $E$ and $F$, $\alpha$ determines where $w_{*}$ will be placed.   

%>>> <<<

As already stated, the crossing of the CC boundary can be realized only for $E \neq 1$ and $F \neq 1$. Moreover, the values $-E$ and $-F$ must be on the opposite sides of the CC boundary. In this setting we are interested to find those $\alpha$ values for which $w_{*}$ is situated in the intervals between $-F$ and $-1$ or between $-E$ and $-1$.
Four possible cases with the corresponding intervals of $\alpha$ are:
\begin{eqnarray}
\label{eq:alphacond}
&F& > -w_{*} > 1 > E \;\;\;\;\;\; (1-F)/(1-E) < \alpha < 0 \, ,  \nonumber \\
&F& > 1 > -w_{*} > E \;\;\;\;\;\; \alpha < (1-F)/(1-E) \, , \nonumber \\
&E& > -w_{*} > 1 > F \;\;\;\;\;\; \alpha < (1-F)/(1-E) \, , \nonumber \\
&E& > 1 > -w_{*} > F \;\;\;\;\;\; (1-F)/(1-E) < \alpha < 0 \, . 
\end{eqnarray}

\begin{figure}
\centerline{\resizebox{0.45\textwidth}{!}{\includegraphics{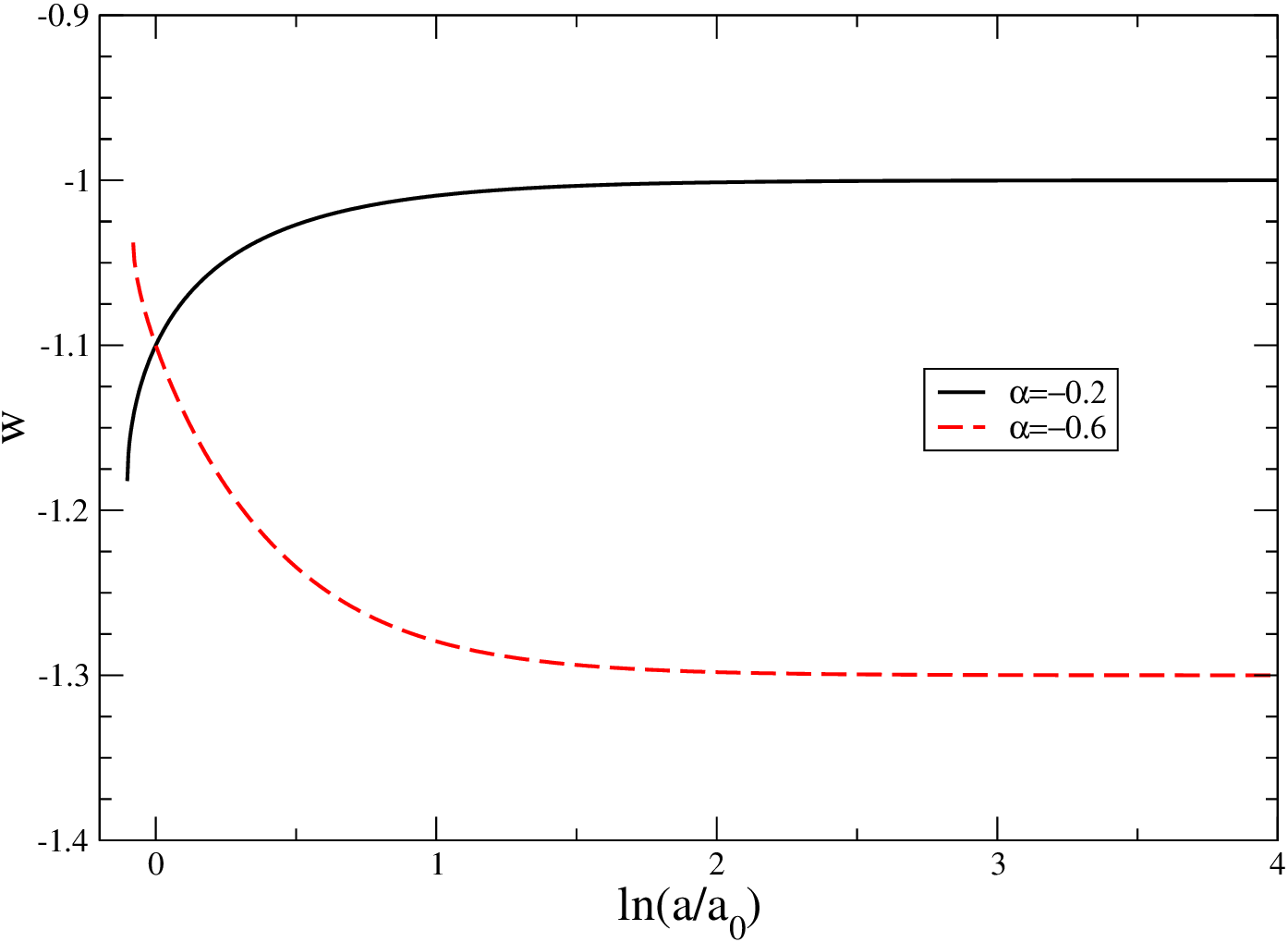}}}
\caption{\label{fig:minus02} The dependence of the parameter $w$ on $a$ for the model (\ref{eq:eosgenmod1}) for two typical values for $\alpha$ in the interval $(1-F)/(1-E) < \alpha < 0$. The parameter values used are $w_{0}=-1.1$, $E=0.6$ and $F=1.3$. The values of the parameter $w_{*}$ are $-1.18$ and $-1.04$ for $\alpha=-0.2$ and $-0.6$, respectivelly, and are within the interval $(-F, -1)$. 
}
\end{figure}

\begin{figure}
\centerline{\resizebox{0.45\textwidth}{!}{\includegraphics{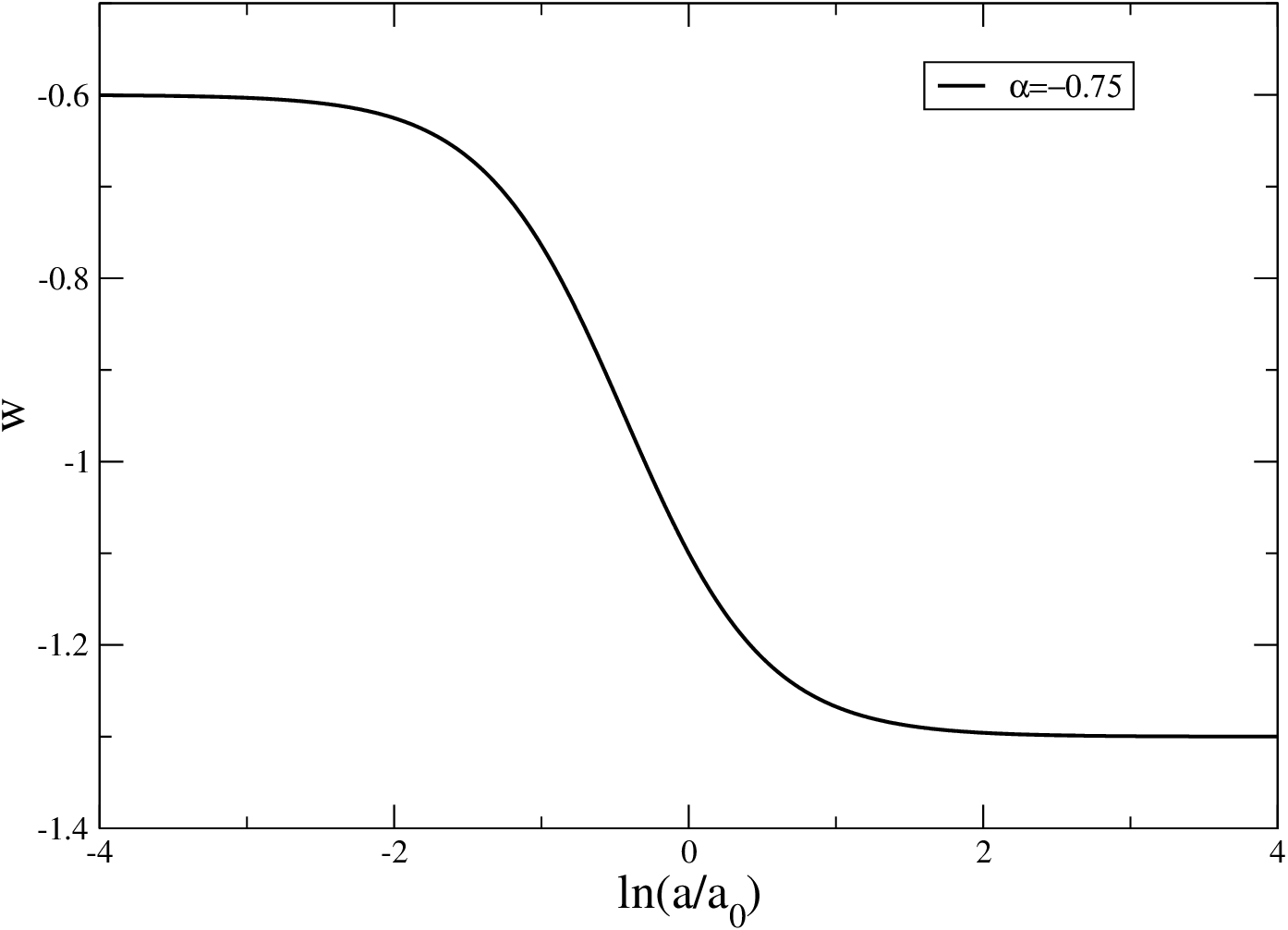}}}
\caption{\label{fig:cross} The parameter $w$ as a function of $a$ for $\alpha=\alpha_{\mathrm cross}=(1-F)/(1-E)$. The parameter values used are $w_{0}=-1.1$, $E=0.6$ and $F=1.3$. The parameter $w_{*}$ equals $-1$. The transition from $w > -1$ to $w < -1$ is nicely demonstrated.  
}
\end{figure}

\begin{figure}
\centerline{\resizebox{0.45\textwidth}{!}{\includegraphics{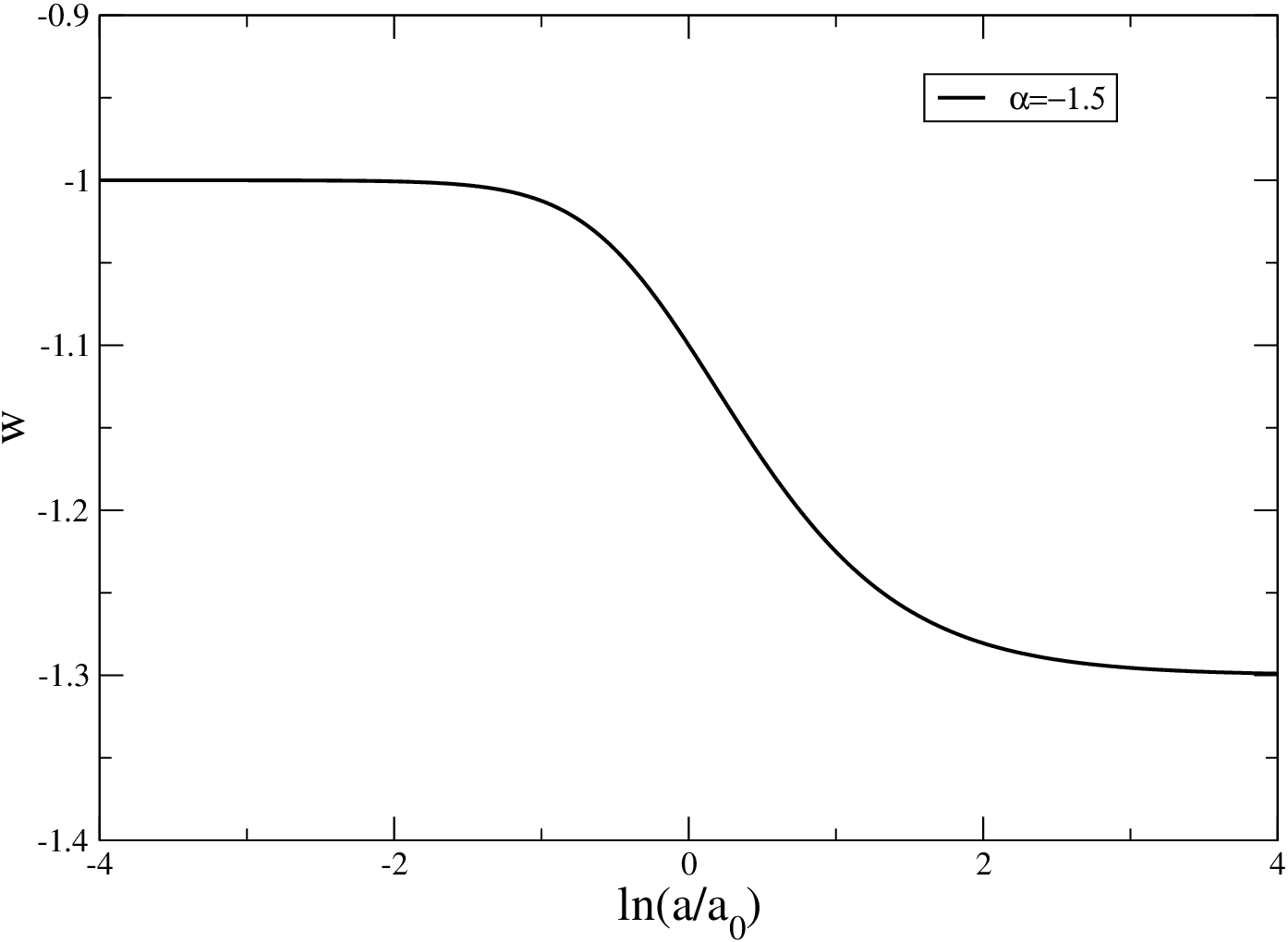}}}
\caption{\label{fig:minus15} The dependence of the parameter $w$ on $a$ for $\alpha < (1-F)/(1-E)$. The parameter values used are $w_{0}=-1.1$, $E=0.6$ and $F=1.3$. For the used value $\alpha=-1.5$, the value of the parameter $w_{*}$ equals $-0.88$ and is situated outside the interval $(-F,-1)$.
}
\end{figure}

Let us consider in detail the behavior of the model as $\alpha$ is decreased from very large positive values toward very negative values for the case $F > -w_{0} > 1 > E$. For $\alpha > 1$, $w \rightarrow -1$ when $a \rightarrow 0$ while $w \rightarrow -F$ when $a \rightarrow \infty$. For $\alpha = 1$, $w=w_{0}$, and for $0 < \alpha < 1$, $w \rightarrow -F$ when $a \rightarrow 0$ and $w \rightarrow -1$ when $a \rightarrow \infty$. The scaling of $w$ with $a$ for these $\alpha$ values is depicted in Fig. \ref{fig:veciod0}. For $\alpha=0$, the EOS becomes the generalized linear EOS studied in detail in \cite{eroshenko} and the boundary at $w=-F$ is removed. For $(1-F)/(1-E) < \alpha < 0$ the parameter $w_{*}$ enters the interval $(-F, -1)$. The form of the dependence of $w$ on $a$ is determined by the relation of $w_{*}$ and $w_{0}$. Namely, the $w(a)$ function has a different form for $w_{*}<w_{0}$ and 
$w_{*} > w_{0}$ as can be seen from Fig. \ref{fig:minus02}. The evolution of the parameter $w$ in this interval of  $\alpha$ implies the existence of the minimal value of the scale factor (the maximal value of the redshift). Clearly, this interval of $\alpha$ is strongly constrained by the cosmological observations. In order to get a very small $a_{*}$ which could possibly be consistent with the observational data, a value very close to $-F$ or $-1$ needs to be chosen for $w_{0}$. However, in this case during the most of the expansion of the universe the behavior of $w$ is practically indistinguishable from the behavior of the dark energy with constant $w=-F$ or $w=-1$ (CC). For $\alpha=(w_{0}+F)/(w_{0}+E)$, we have $a_{*}=a_{0}$ and $w_{*}=w_{0}$ which is observationally completely excluded. 
Clearly, this interval is of small observational importance and it is discussed here for the sake of the exposition completeness. For $\alpha=\alpha_{\mathrm cross}=(1-F)/(1-E)$ the transition between the quintessence and the phantom regimes of dark energy occurs, as displayed in Fig. \ref{fig:cross}. Finally, for $\alpha < \alpha_{\mathrm cross}$, $w \rightarrow -1$ when $a \rightarrow 0$ and $w \rightarrow -F$ when $a \rightarrow \infty$, as shown in Fig. \ref{fig:minus15}. 

Next we describe the behavior of the model in the parameter regime $F >1 > -w_{0} > E$ considering $\alpha$ values from very large positive to very negative ones. For $\alpha > 1$, $w \rightarrow -1$ when $a \rightarrow 0$ and $w \rightarrow -E$ when $a \rightarrow \infty$. When $\alpha=1$, $w=w_{0}$ and for  $(1-F)/(1-E) < \alpha < 1$, $w \rightarrow -E$ when $a \rightarrow 0$ and $w \rightarrow -1$ when $a \rightarrow \infty$. At $\alpha=\alpha_{\mathrm cross} = (1-F)/(1-E)$ the parameter $w$ transits from $-E$ to $-F$ with the expansion. For $(w_{0}+F)/(w_{0}+E) < \alpha < (1-F)/(1-E)$,  $w \rightarrow -E$ when $a \rightarrow 0$ and $w \rightarrow w_{*}$ when $a \rightarrow a_{*}>a_{0}$. For $\alpha=(w_{0}+F)/(w_{0}+E)$, we have $w_{*}=w_{0}$ and $a_{*}=a_{0}$ which is also observationally excluded. Finally, for $\alpha <(w_{0}+F)/(w_{0}+E)$, $w \rightarrow -1$ when $a \rightarrow 0$ and $w \rightarrow w_{*}$ when $a \rightarrow a_{*}>a_{0}$. It is interesting to note that some of the cases for $\alpha < (1-F)/(1-E)$, characterized by the maximal value of the scale factor reached in the future, might be interesting in the studies of future singularities of the expansion of the universe \cite{odin1,PRDja,Barrow,sudden}. The behavior of the model in two remaining interesting cases $E > -w_{0} > 1 > F$ and $E >1 > -w_{0} > F$ may be obtained from the results displayed above since the model (\ref{eq:evolw1}) is symmetric with respect to transformations $\alpha \leftrightarrow 1/\alpha$ and $E \leftrightarrow F$.

More general models, characterized by the transition between the quintessence and phantom regimes, can be explicitly constructed. Namely, the models of dark energy density given by   
\begin{equation}
\label{eq:densitymod2}
\rho = \left( C_{1} \left( \frac{a}{a_{0}} \right)^{-3(1+\gamma)/b} + 
C_{2} \left( \frac{a}{a_{0}} \right)^{-3(1+\eta)/b} \right)^{b} \, , 
\end{equation}
characterized by an additional parameter $b$, exhibit the crossing of the CC boundary. The dark energy pressure becomes % 
\begin{equation}
\label{eq:pressuremod2}
p \rho^{(1-b)/b} = \gamma C_{1} \left( \frac{a}{a_{0}} \right)^{-3(1+\gamma)/b} + 
\eta C_{2} \left( \frac{a}{a_{0}} \right)^{-3(1+\eta)/b} \, ,
\end{equation}
and the parameter of dark energy EOS is
\begin{equation}
\label{eq:wformod2}
w = \frac{\gamma + \eta \frac{\gamma - w_{0}}{w_{0} - \eta} \left( \frac{a}{a_{0}} 
\right)^{3(\gamma - \eta)/b}}{1 + \frac{\gamma - w_{0}}{w_{0} - \eta} \left( \frac{a}{a_{0}} 
\right)^{3(\gamma - \eta)/b}} \, .
\end{equation}

\begin{figure}
\centerline{\resizebox{0.45\textwidth}{!}{\includegraphics{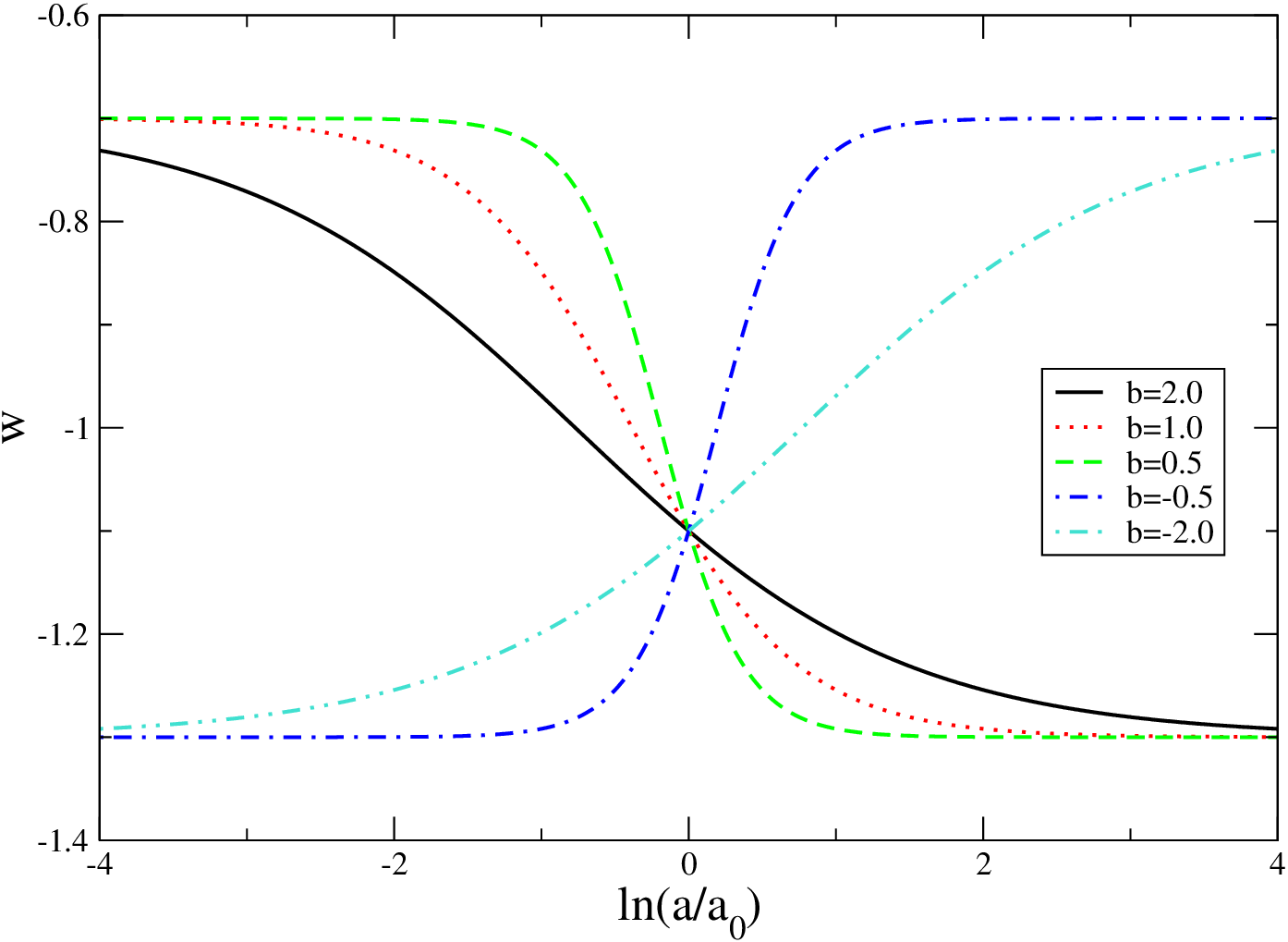}}}
\caption{\label{fig:mod2} The dependence of the parameter of the dark energy EOS $w$ on the scale factor of the universe $a$ for the model (\ref{eq:densitymod2}). The positive values of the parameter $b$ describe the transition from $\gamma$ to $\eta$ with the expansion of the universe, whereas the negative values for $b$ describe the transition from $\eta$ to $\gamma$ as the universe expands. The parameter values used are $w_{0}=-1.1$, $\gamma=-0.7$ and $\eta=-1.3$.   
}
\end{figure}

The behavior of the model depending on parameter $b$ is depicted in Fig. \ref{fig:mod2}. The additional new feature obtained by the introduction of the parameter $b$ is the possibility to describe the transitions in both directions with respect to the asymptotic values of the parameter $w$, see Fig. \ref{fig:mod2}. Using the procedure applied to the model (\ref{eq:denmod1}), we obtain the dark energy EOS
\begin{equation}
\label{eq:eosdetmod2}
\frac{p -\eta \rho}{(\gamma-\eta) C_{1}} =
\rho^{((1-b)(\gamma - \eta))/(b(1+\eta))}
\left( \frac{\gamma \rho - p}{(\gamma-\eta) C_{2}} \right)^{(1+\gamma)/(1+\eta)}
\, .
\end{equation}
This dark energy EOS is, as in the case of the model (\ref{eq:denmod1}), defined {\em implicitly}. A natural generalization of the model (\ref{eq:densitymod2}) is given by an EOS
\begin{equation}
\label{eq:eosmodel2}
A \rho + B p = (C \rho + D p)^{\alpha} (M \rho + N p)^{\beta} \, ,
\end{equation}
where $A$, $B$, $C$, $D$, $M$, $N$, $\alpha$, and $\beta$ are real parameters. The dark energy density has the following dependence on $w$:
\begin{equation}
\label{eq:rhomod2}
\rho=\left( \frac{(C + D w)^{\alpha} (M + N w)^{\beta}}{A + B w} \right)^{1/(1-\alpha-\beta)} \, .
\end{equation}
Here we assume that $\alpha+\beta \neq 1$. In the case when $\alpha+\beta = 1$ the equation (\ref{eq:rhomod2}) becomes an algebraical equation for the parameter $w$. The model is then a dark energy model with a constant parameter of the EOS obtained by solving the aforementioned algebraical equation.    
Inserting the expression for the energy density into (\ref{eq:bianchi}) we obtain the equation of evolution of the parameter $w$ with the scale factor $a$: 
\begin{eqnarray}
\label{eq:wmod2}
&&\left( \frac{\alpha D}{C+D w}+\frac{\beta N}{M+N w}-
\frac{B}{A+B w} \right) \frac{dw}{1+w} \nonumber \\
&=& 3(\alpha+\beta - 1) \frac{da}{a} \, .
\end{eqnarray}
The solution of the equation given above for the most interesting case $A \neq B$, $C \neq D$, and $M \neq M$ gives the expression for $w$ in the closed form as
\begin{eqnarray}
\label{eq:solwmod2}
& & \left| \frac{C+D w}{C+D w_{0}} \right|^{-\alpha D/(C-D)}
\left| \frac{M+N w}{M+N w_{0}} \right|^{-\beta N/(M-N)} \nonumber \\
&\times& \left| \frac{1+w}{1+w_{0}} \right|^{\alpha D/(C-D)+\beta N/(M-N)-B/(A-B)} \nonumber \\
&\times& \left| \frac{A+B w}{A + B w_{0}} \right|^{B/(A-B)} 
= \left( \frac{a}{a_{0}} \right)^{3(\alpha+\beta-1)} \, . 
\end{eqnarray}
This expression shows that it is possible to remove the CC boundary if a special relation among the model parameters is imposed (which is essentially the requirement that the exponent of the $|1+w|$ term vanishes).   

The models considered so far were either constructed to exhibit the crossing of the CC boundary or obtained as the generalizations of the explicitly constructed models. Therefore, in the case of models obtained as generalizations of the constructed models, it is certain that at least some combinations of model parameters yield the transition. Based on the insight obtained in the study of the dark energy models discussed so far, it is possible to show that a dark energy model with a highly nontrivial implicitly defined EOS also exhibits the crossing of the CC boundary for some values of model parameters. Namely, let us consider the EOS
\begin{equation}
\label{eq:ton}
A \rho^{2n+1} + B p^{2n+1} = (C \rho^{2n+1} + D p^{2n+1})^{\alpha} \, ,
\end{equation}
where $n \ge 0$. The evolution equation for the EOS parameter $w$ is
\begin{equation}
\label{eq:wforn}
\frac{w^{2n+1}+(\alpha E -F)/(\alpha-1)}{(F+w^{2n+1})(E+w^{2n+1})} 
\frac{w^{2n}}{1+w} dw = 3 \frac{da}{a} \, .
\end{equation}
The conditions for the occurrence of the transition are met when $\alpha=(1-F)/(1-E)$, i.e. $(\alpha E -F)/(\alpha-1)=1$. In this case we have a cancellation of the $(w+1)$ terms between the $w^{2n+1}+1$ term in the numerator and the $w+1$ term in the denominator. Namely,
\begin{equation}
\label{eq:cancel}
\frac{w^{2n+1}+1}{w+1} = \xi(w) = \sum_{l=0}^{2n} (-w)^{l} \, .
\end{equation}
Here the function $\xi(w)$ has no real roots and therefore, no additional  analogue of $w_{*}$ can appear. The evolution equation becomes
\begin{equation}
\label{eq:cancel2}
\frac{\xi(w)}{(F+w^{2n+1})(E+w^{2n+1})} 
w^{2n} dw = 3 \frac{da}{a} \, ,
\end{equation}
and it describes the smooth transitions of $w$ between $-E^{1/(2n+1)}$ and $-F^{1/(2n+1)}$. Therefore, the dark energy model with the implicitly defined EOS (\ref{eq:ton}) is capable of describing the dark energy transition between quintessence and phantom regimes. 

The main result of this paper is the demonstration that it is possible to describe the crossing of the CC boundary using the dark energy EOS only. The important and new difference with respect to similar studies until now is that the equations of state considered here are defined {\em implicitly}. The understanding of the dark energy EOS as parametrically defined (like a pair of quantities $(\rho(t),p(t))$) certainly opens much larger possibilities in describing the dark energy evolution and properties. The transition between $w>-1$ and $w<-1$ regimes of the dark energy behavior is one of aspects that can be described in terms of implicitly defined dark energy EOS. 

Another aspect of the crossing of the CC boundary in the framework of the implicitly defined dark energy EOS is related to the conditions that must be fulfilled for the transition to take place. In all studied models which have not been explicitly constructed to yield the transition, i.e. in generalized models, one parameter had to have a special value determined by some function of remaining parameters. A deviation from this special value prevents the occurrence of the transition. In other words, if the parametric space of the model is $D$ dimensional, the set of parameter values which lead to the transition is $D-1$ dimensional (i.e. determined by $D-1$ parameters). Clearly, if no additional mechanism selected parameter values that correspond to the transition, i.e. if all parameter combinations were equally likely, it could be said that the transition of the CC boundary in the so far studied dark energy models with the implicitly defined EOS is unlikely.
However, only a couple of dark energy models with the implicitly defined EOS have been studied in this paper. It remains to be seen whether a more suitable implicitly defined dark energy EOS can make the transition more likely.    
%samo su neke EOS proucavane, ima i jos.
It is important to stress that, although it was not the main aim of this paper, the dark energy models with the implicitly defined EOS studied in this paper exhibit some additional interesting features and effects. Namely, the generalized models like (\ref{eq:eosgenmod1}) exhibit the smooth transitions between any two values which are both less or bigger than -1. These transitions are, in the sense of the discussion given in the preceding paragraph, likely. Furthermore, the model (\ref{eq:eosgenmod1}) exhibits an intriguing behavior around $w_{*}$ which could be interesting in the study of (sudden) future singularities.  

%\begin{figure}
%\centerline{\resizebox{0.45\textwidth}{!}{\includegraphics{scalea1.eps}}}
%\caption{\label{fig:scalea1}  The time evolution of the scale factor of the universe for
%$\Omega_{d,0}=0.7$, $\Omega_{m,0}=0.3$,
%$\tilde{A} = 1$ and four typical values of the parameter $\alpha$: $\alpha = -1,0.5,0.75,2$.
%For $\alpha=2$, the scale factor reaches the finite value at the singularity while for $\alpha=0.75$, the scale %factor of the universe diverges at singularity.}
%\end{figure}
%\vspace{1cm}
%{\em sudden future singularities}

In conclusion, the dark energy transition between quintessence and phantom regimes is studied in terms of the EOS. The dark energy EOS for models explicitly constructed to exhibit the transition is found to be implicitly defined. Several generalized implicitly defined dark energy EOS are studied to investigate the conditions which are necessary for the occurrence of the transition. Within the generalized models studied in this paper it is found that special conditions need to be satisfied in order to have a transition.
The mechanism behind the transition within the generalized models is related to the cancellation of the contribution from the CC boundary. It is important to emphasize that, once it has been shown that the model with the implicitly defined EOS may describe the crossing of the CC boundary, it is reasonable to investigate implicitly defined EOS which could describe other interesting effects, such as a transient phantom phase. It is also of interest to investigate other classes of dark energy models with the implicitly defined EOS to find out whether the special conditions required for the transition could be relaxed.   

{\bf Acknowledgments.} The author acknowledges the support of the Secretar\'{\i}a de Estado de Universidades e Investigaci\'{o}n of the Ministerio de Educaci\'{o}n y Ciencia of Spain within the program ``Ayudas para movilidad de Profesores de Universidad e Investigadores espa\~{n}oles y extranjeros". This work has been supported in part by MEC and FEDER under project 2004-04582-C02-01 and by the Dep. de Recerca de la Generalitat de Catalunya under contract CIRIT GC 2001SGR-00065. The author would like to thank the Departament E.C.M. of the Universitat de Barcelona for the hospitality.

\end{document}